\documentclass[pra,a4paper,twocolumn,floatfix]{revtex4}

\usepackage{graphicx}

\begin{document}
\title{Collective dynamics and expansion of a Bose-Einstein condensate \\
in a random potential} \author{M. Modugno} 
%\email{modugno@fi.infn.it}
\affiliation{Dipartimento di Matematica Applicata, Universit\`a di
Firenze, I-50139 Firenze, Italy} 
\affiliation{LENS and INFM, Universit\`a di Firenze,
I-50019 Sesto Fiorentino (FI), Italy} \affiliation{BEC-INFM Center,
Universit\`a di Trento, I-38050 Povo (TN), Italy}

\date{\today}

\begin{abstract}    

We investigate the dynamics of a Bose-Einstein condensate in the
presence of a random potential created by optical speckles.  We
first consider the effect of a weak disorder on the dipole and
quadrupole collective oscillations, finding uncorrelated frequency
shifts of the two modes with respect to the pure harmonic
case. This behaviour, predicted by a  sum rules
approach, is confirmed by the numerical solution of the
Gross-Pitaevskii equation.  Then we analyze the role of disorder
on the one-dimensional expansion in an optical guide, discussing
possible localization effects.  Our theoretical analysis provides
a useful insight into the recent experiments performed at LENS
[J. Lye \textit{et al.}, Phys. Rev. Lett. \textbf{95}, 070401 (2005); 
C. Fort \textit{et al.}, cond-mat/0507144].

\end{abstract}

\maketitle

%-------------------------------------------
\section{Introduction}
%-------------------------------------------

The investigation of Bose-Einstein condensates (BECs) in the presence
of disorder is rapidly becoming a central topic in ultracold atom
physics \cite{roth,damski,castin,lye,fort,clement,ertmer,paul}.
Bosonic systems in disordered potentials have been extensively investigated
in the recent past, both experimentally and theoretically
\cite{disorder}. Experiments with superfluid $^4$He in porous
materials have demonstrated the suppression of superfluid transport
and the critical behavior at the phase transition in presence of
disorder \cite{helium}.  From the theoretical point of view a rich
variety of phenomena is expected to occur in these systems, among
which the most fascinating are Anderson localization, initially
proposed in the context of electron transport in disordered solids
\cite{anderson}, and later predicted and observed for 
non-interacting wave phenomena such as light \cite{w2,w3}, and the quantum
transition to the Bose glass phase that originates from the interplay
of interactions and disorder \cite{boseglass}.

The demonstrated capability of using BECs as versatile tools to
revisit condensed matter physics \cite{latticereview}, as for example
the transition from superfluid to Mott insulator \cite{greiner},
suggests that these are also promising tools to engineer disordered quantum
systems \cite{roth,damski,castin}. Recently, effects of disorder
created by a laser speckle have been observed on the dynamics of a
BEC, including uncorrelated shifts of the quadrupole and dipole modes
\cite{lye} and localization phenomena during the expansion in a
one-dimensional (1D) waveguide \cite{fort,clement}.
Effects of disorder have also been observed for BECs in microtraps 
as a consequence of intrinsic defects in the fabrication of the microchip
\cite{fragmentation, Fortagh}.

In Refs. \cite{lye,fort} we have shown that the main features observed
in that experiments can be explained within the Gross-Pitaevskii (GP)
theory. In this paper we report a detailed analysis and discussion of
the theoretical approach used, comparing the effects of different
kinds of random potentials.  We also make a systematic comparison
with the case of a periodic lattice with spacing of the order of the
length scale of disorder. This helps to discriminate the effects due
to the particular realization of the random potential from those that
are intimately connected to the disorder.

We show that in the presence of a weak disorder the dipole and
quadrupole modes of a harmonically trapped condensate are undamped in
the small amplitude regime, whereas a superfluid breakdown may occur
for larger oscillations. In the first case the two modes are
characterized by uncorrelated frequency shifts, both in sign and
amplitude, that depend on the particular realization of the perturbing
potential.  The average features however do not depend crucially on
the particular kind of disorder, but still
evidence significant differences with the periodic case.  We also
show that the localization effects observed during the expansion in a
1D waveguide are mainly due to a classical trapping into
single wells or between barriers of the random potential. The
qualitative behavior in this case is very similar to that of a
periodic system.

The paper is organized as follows: we start in Sect. \ref{sec:system}
by describing the system and the various kinds of disorder considered.
Then in Sect. \ref{sec:collective} we discuss the effect of the random
potential on the dipole and quadrupole collective oscillations of the
system by means of a sum rules approach and the direct solution of the
GP equation.  Next, in Sect. \ref{sec:expansion} we address the role
of disorder on the BEC expansion in a 1D waveguide by analyzing the
results of the GP calculations in terms of the quantum behaviour of a
single defect (well/barrier) of the potential.  A detailed description
on the numerical characterization of the random potentials is reported
in the Appendix.

%-------------------------------------------
\section{Description of the system}
\label{sec:system}
%-------------------------------------------

In this paper we will consider the case of an elongated condensate
confined in a cylindrically symmetric harmonic potential
\begin{equation} 
V_{ho}(r_\perp,z) =\frac12 m\omega_\perp^2
r_\perp + \frac12 m\omega_z^2 z^2
\end{equation}
and subjected to an additional random potential $V_R(z)$ along the
axial direction. The latter is characterized by the correlation length
$l_c$ and the amplitude $V_0$, and can be written as $V_R(z)=V_0 v(z)$
with the distribution of intensities of $v(z)$ being normalized to
unit standard deviation.  Here we will address three kinds of disorder:
two corresponding to a laser speckle potential $\pm v_S(z)$, where the
$\pm$ indicates whether it is red or blue detuned (that is, the
potential can be attractive or repulsive), and another generated by a
gaussian random potential $v_G(z)$. A detailed description on how the
 potentials are constructed and characterized is reported in
Appendix \ref{sec:random}.

\begin{figure}
\centerline{\includegraphics[height=0.95\columnwidth,clip=,angle=-90]{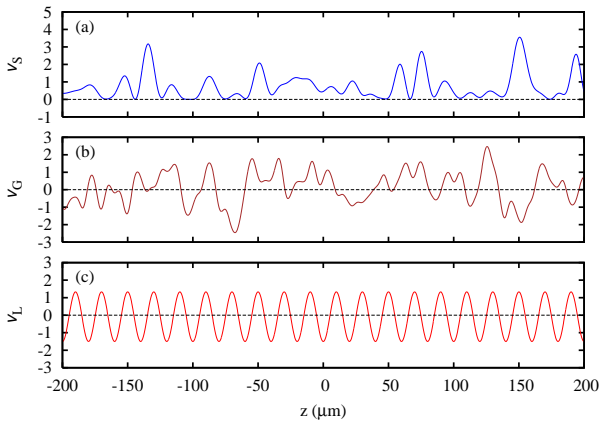}}
\caption{(Color online) Typical shape of the potentials considered in
this paper: (a) (blue detuned) speckles (red speckles would have the
same shape but reversed sign), (b) gaussian random, (c) periodic. In
all the cases the correlation length is $l_c=10\mu$m.}
\label{fig:pot}
\end{figure}

In some cases it will also be useful to compare the effect of disorder
to the case of a periodic lattice $V_L(z)=V_0 v_L(z)$. For this
purpose a suitable choice is a sinusoidal potential
$v_L(z)=2\sqrt{2}\sin^2(\pi z/2l_c)$ with intensity normalized as
before and whose wavevector is chosen to match the correlation length
of the random potential, as discussed in Appendix \ref{sec:random}.
The typical shape of the various potentials for a correlation length
$l_c=10\mu$m is depicted in Fig \ref{fig:pot}.

%-------------------------------------------
\section{collective excitations}
\label{sec:collective}
%-------------------------------------------

Let us start by discussing the effect of a random potential on the
collective excitations of the system, considering in particular the
dipole and quadrupole modes. First we will consider the regime of weak
disorder and small amplitude oscillations, by comparing the prediction
of sum rules with the numerical solution of the GP
equation, analyzing in more details the theoretical description of the
experiments reported in \cite{lye}.  At the end of the section we
will also briefly discuss the possibility of a superfluidity
breakdown that may occur for larger amplitude oscillations.

%-------------------------------------------
\subsection{Sum rules approach}
%-------------------------------------------

A powerful tool for characterizing the collective frequencies of the
system is the sum rules approach \cite{stringari,kimura}. Within this
approach an upper bound for the frequencies of the low lying
collective excitations of a many-body system is given by
\begin{equation}
\omega^2=\frac{1}{\hbar^2}\frac{m_3}{m_1}
\end{equation}
where the moments $m_i$ are defined via the following commutators
\begin{eqnarray}
m_1&=&\langle\left[F,[H,F]\right]\rangle \\
m_3&=&\langle\left[[F,H],[H,[H,F]]\right]\rangle
\end{eqnarray}
between the many-body hamiltonian $H$
and a suitable excitation operator $F$ that is chosen as follows
\begin{eqnarray}
F_D &=& z\\ F_Q &=& r_\perp^2 -\alpha^2 z^2
\end{eqnarray}
$\alpha$ being a variational parameter. 
In our case the hamiltonian can be written as
\begin{equation}
\label{eq:hamilt} H =
\sum_{i=1}^{N}\left[\frac{\hat{\textbf{p}}_i^2}{2m} + V_{ho}+ V_R
+g\sum_{j=1}^{i-1} \delta(\textbf{x}_j-\textbf{x}_i)\right]
\end{equation}
where the interaction strength 
$g$ is related to the interatomic
scattering length $a$ by $g=4\pi\hbar^2a/m$, $m$ being the atomic mass.

In case of the harmonic potential $V_{ho}$ alone (unperturbed case)
the dipole and quadrupole collective frequencies have the well known
expressions $\omega_D=\omega_z$ for the dipole mode along $z$,  
and $\omega_Q=\sqrt{5/2}\,\omega_z$ for the quadrupole mode in  
case of an elongated condensate in the large $N$ Thomas-Fermi (TF) limit.

Let us now discuss the effect of a shallow random potential $V_R(z)$.
Treating the $V_R$ as a small perturbation, and writing
$\omega^2=\omega_0^2+\delta$ we get
\begin{eqnarray}
\delta_D&\simeq&\frac{1}{m}\langle\partial_z^2V_R\rangle_0 
\label{eq:dipole_shift}\\
\delta_Q&\simeq&\frac{1}{m}\frac{\langle
z\partial_zV_R+z^2\partial_z^2V_R\rangle_0}{\langle z^2\rangle_0}
\label{eq:quadrupole_shift}
\end{eqnarray}
where the averages $\langle \cdot\cdot\rangle_0$ are calculated on the
unperturbed ground state, and the second line is obtained assuming a
strongly elongated condensate. 

The above equations imply that in general the shifts of two frequencies are 
uncorrelated and depend on the particular shape of the perturbing
potential and on its relative position respect to the harmonic
potential.

\begin{figure}
\centerline{\includegraphics[width=0.95\columnwidth,clip=]{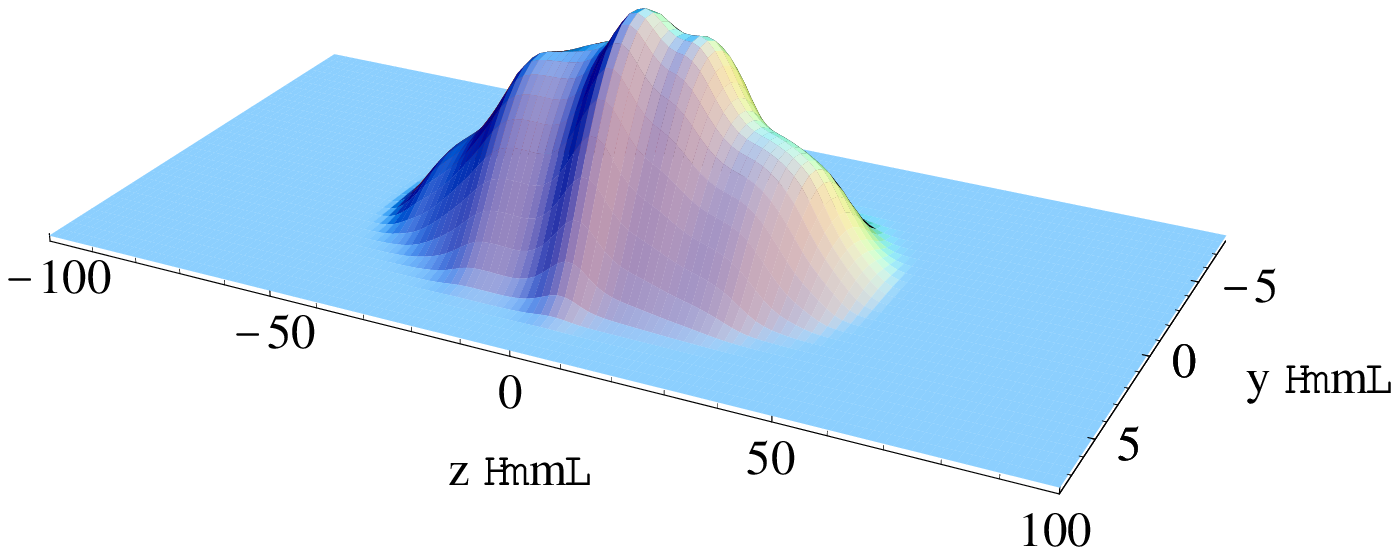}}
\centerline{\includegraphics[height=0.95\columnwidth,clip=,angle=-90]{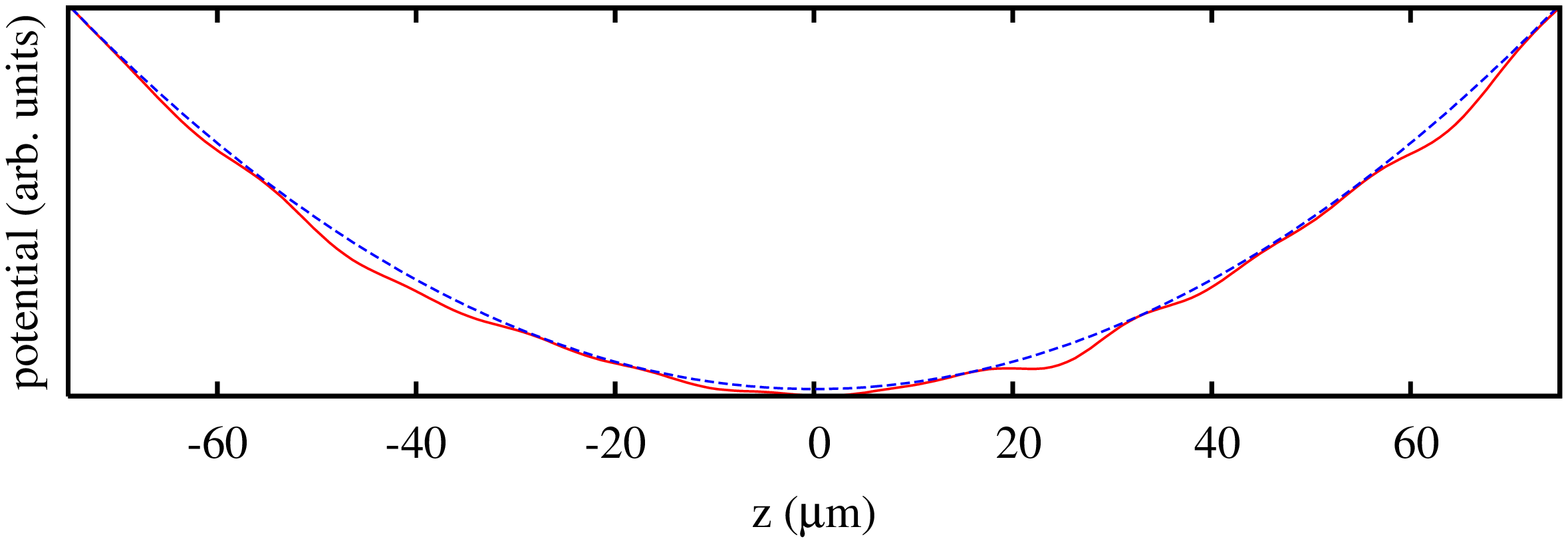}}
\caption{(Color online) (a) Column density of a typical ground state
configuration in the presence of a red-detuned speckle potential
($V_0=2.5\,\hbar\omega_z$).
(b) Solid line: profile of the combined potential
$V_R$ + $V_{ho}$ (dashed line).}
\label{fig:ground}
\end{figure}
To show how this works in a particular example we consider here the
typical parameters of the LENS experiment in \cite{lye}: frequencies
$\omega_z=2\pi\times 9$ Hz and $\omega_\perp=2\pi\times 90$ Hz, total
number of atoms $N=1\times 10^5$, $V_0=2.5\;\hbar\omega_z$ and
$l_c=10\mu$m. A picture of the total potential resulting from these
parameters and of the corresponding ground state is shown in
Fig.~\ref{fig:ground}.

\begin{figure}
\includegraphics[height=0.95\columnwidth,clip=,angle=-90]{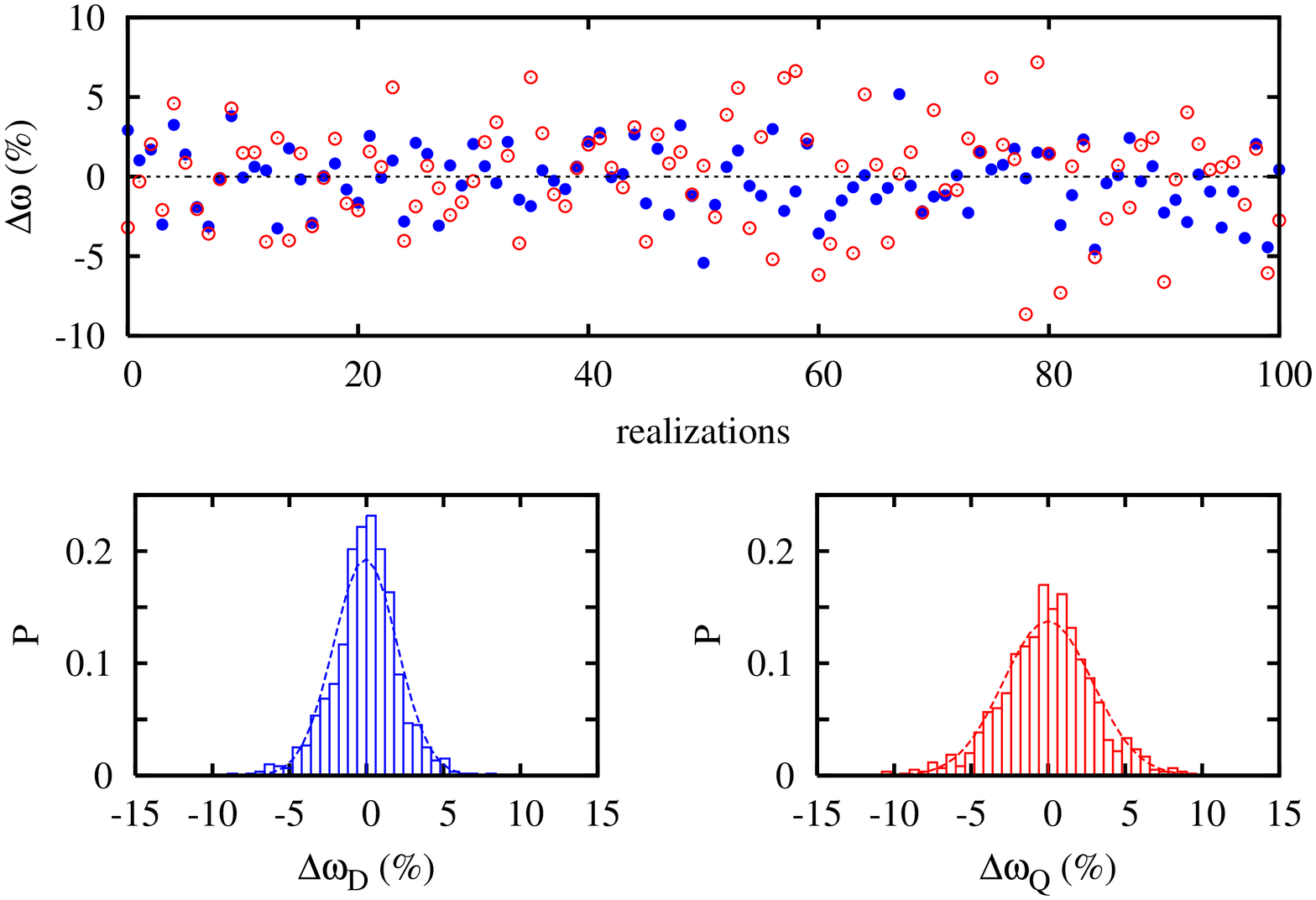}
\caption{(Color online) Top: dipole (filled circles) and quadrupole 
(empty circles) frequency shifts
for 100 different realizations of the speckle potential as obtained
from the sum rules. Bottom: their probability distribution $P$ for
1000 realizations (left:dipole, right:quadrupole); the dashed lines are 
a gaussian fit.}
\label{fig:sumrules}
\includegraphics[height=0.95\columnwidth,clip=,angle=-90]{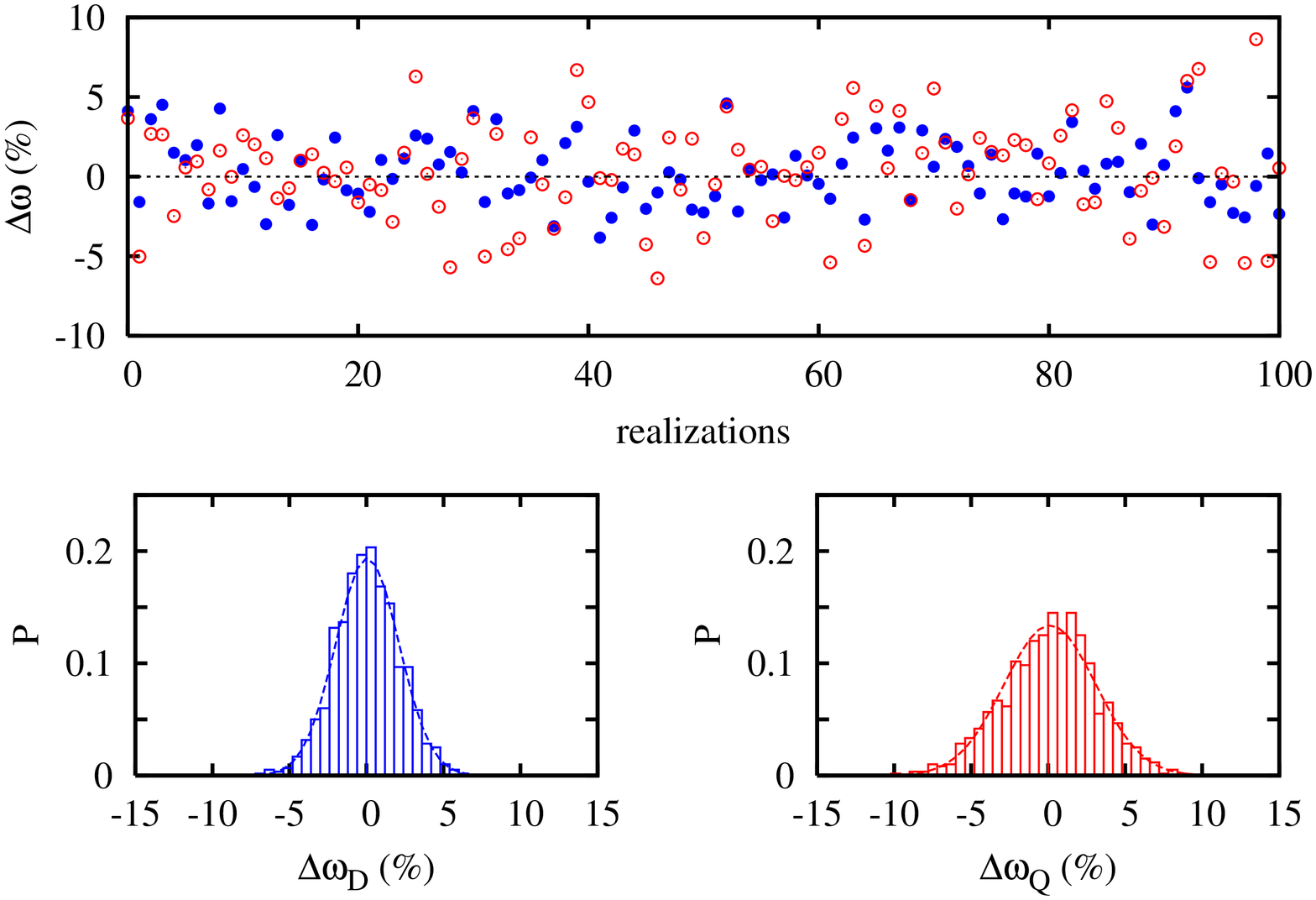}
\caption{(Color online) Top: dipole (filled circles) and quadrupole 
(empty circles) frequency shifts
for 100 different realizations of the gaussian random potential as obtained
from the sum rules. Bottom: their probability distribution $P$ for
1000 realizations (left:dipole, right:quadrupole); the dashed lines are 
a gaussian fit.}
\label{fig:sumrules_g}
\end{figure}

In Fig.~\ref{fig:sumrules} we show the frequency shifts
$\Delta\omega\equiv\omega-\omega_0\simeq\delta/2\omega_0$ and their
statistical distributions respectively for 100 and 1000 different 
realizations of the speckle potential. 
The picture shows that the dipole and quadrupole
shifts are uncorrelated, in contrast to what happens in case of a pure
harmonic potential or in the presence of a periodic potential
\cite{kramer}.  This behavior does not depend on whether the speckles
are red or blue detuned (according to Eqs. \ref{eq:dipole_shift} and
\ref{eq:quadrupole_shift} this corresponds to a change of sign, and
therefore the statistical properties in Fig.~\ref{fig:sumrules} remain
unchanged). We have also verified that the behavior is essentially the same
also in case of a gaussian disorder, see Fig.~\ref{fig:sumrules_g}.

\begin{figure}
\includegraphics[height=0.95\columnwidth,clip=,angle=-90]{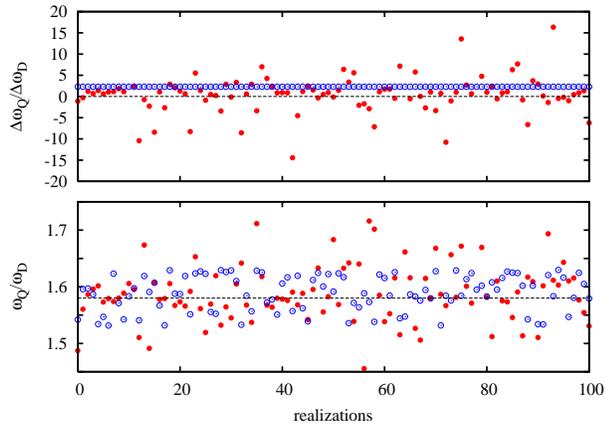}
\caption{(Color online) Ratios between quadrupole and dipole shifts
(top) and the corresponding frequencies (bottom) for speckle (filled
circles) and periodic (empty circles) potentials with $l_c=10\mu$m.
Note that although the shifts for the periodic case are correlated in
sign the two frequencies are uncorrelated and depend on the actual
position of the condensate in the periodic potential.}
\label{fig:ratios}
\end{figure}

It is useful also to comment on the behaviour in the presence of a
periodic potential. When the wavelength $\pi/q=2l_c$ of the potential
is much smaller that the axial extent of the condensate one can apply
the Bloch picture and resort to the effective mass approximation.  As
stated above, this yields the same renormalization for both the dipole
and quadrupole frequencies $\omega=\sqrt{m/m^*}\omega_0$, $m^*$ being
the effective mass \cite{kramer}.  Differently, in the case considered
here ($l_c=10\mu$m) the condensate extends over only few wells of the
periodic potential and the Bloch picture cannot be applied. In this
case the sum rules approach predicts a sign correlated shift for the
two frequencies, whose magnitude however still depends on the relative
position between the condensate and the periodic potential. Therefore,
although $\Delta\omega_Q/\Delta\omega_D\sim$ constant, the fact that
the ratio $\omega_Q/\omega_D$ depends at first order on the difference
$\Delta\omega_Q-\Delta\omega_D$ eventually yields an uncorrelated
renormalization of the two frequencies (see Fig.~\ref{fig:ratios}).

%-------------------------------------------
\subsection{GP dynamics}
\label{sec:gpe}
%-------------------------------------------

The prediction of the sum rules can be directly compared with the
solution of the GP equation \cite{bec_review}
\begin{equation}
i\hbar\partial_t\psi=\left[-\frac{\hbar^2}{2m}\nabla^2 + V_{ho} + V_R
+g|\psi|^2\right]\psi
\end{equation}
by exciting the collective modes with a sudden displacement of the
harmonic trap (for the dipole) or a change of the axial trapping
frequency (for the quadrupole).

\begin{figure}
\includegraphics[height=0.95\columnwidth,clip=,angle=-90]{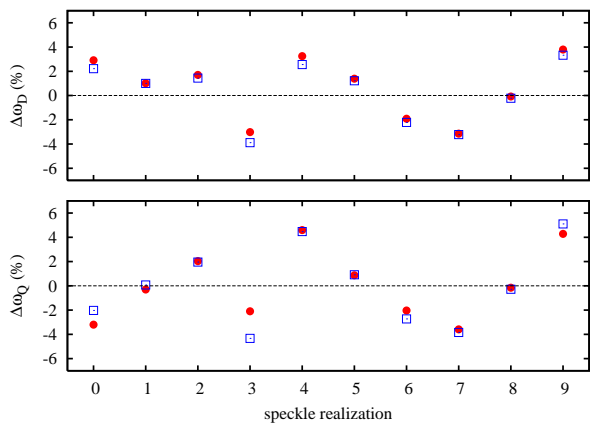}
\caption{(Color online) Dipole (top) and quadrupole (bottom) frequency
shifts as obtained from the GP equation (squares) and compared with the sum
rules predictions (circles), for ten different speckle realizations.
The agreement is remarkable \cite{nota1}.}
\label{fig:gpe}
\end{figure}

The results from some sample realization of the random potential are
shown in Fig.~\ref{fig:gpe} (they correspond to the first ten
realizations in Fig. \ref{fig:sumrules}). The dipole oscillations are
induced after a displacement $\Delta z = 5\,\mu$m of the harmonic
potential, corresponding to an oscillation of the order of $10\%$ of
the axial size of the condensate. For the quadrupole mode, an
oscillation of the same order of magnitude is obtained by releasing
the condensate from a tighter trap of axial frequency
$\omega_z'=1.1\omega_z$.

In this regime of \textsl{small amplitude oscillations} the solution
of the GP equation shows that the condensate oscillates coherently with no
appreciable damping on a timescale of several oscillations. The
corresponding frequencies show a remarkable agreement with the sum
rules predictions, regarding both the sign and order of magnitude of
the shift, as shown in Fig.~\ref{fig:gpe} \cite{nota1}. 
As mentioned above, these
features have been observed in the experiment reported in \cite{lye}.

\begin{figure}
\includegraphics[height=0.95\columnwidth,clip=,angle=-90]{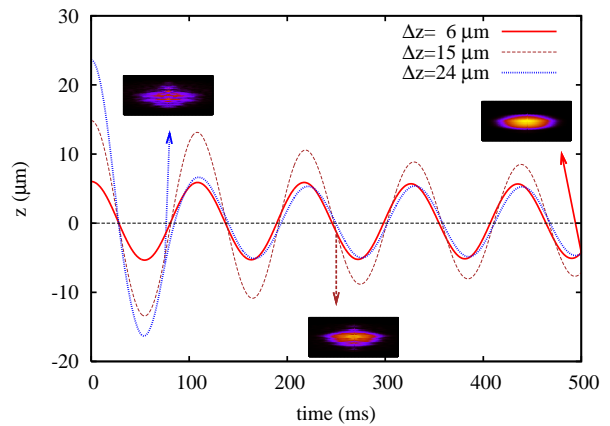}
\caption{(Color online) Dipole oscillations in the speckle potential 
($V_0=2.5\,\hbar\omega_z$)
for three different displacements 
$\Delta z$ of the harmonic potential ($\Delta z =
6,15,24\,\mu$m).  The insets show a  density plot for
$\Delta z = 24\mu$m at $t=75$ ms (left), $\Delta z = 15\mu$m at
$t=250$ ms (center), and $\Delta z = 6\mu$m at $t=500$ ms (right).}
\label{fig:large}
\end{figure}

We have also explored the behavior of the system for \textsl{larger
amplitudes} in case of dipole oscillations, as shown in
Fig.~\ref{fig:large}.  As the amplitude is increased the frequency
shift reduces owing to the fact that the condensate experiences outer
regions of the harmonic potential where the effect of the random
potential is negligible, yielding an average frequency that is closer
to the unperturbed value. However, as the center of mass velocity
increases, the oscillations also gets damped due to the presence of
the speckle potential that acts as an external perturbation or
roughness of the medium. In this regime the condensate develops short
wavelength density modulation that may eventually lead to a breakdown
of the superfluid flow, as shown in the left and center insets in
Fig.~\ref{fig:large} (respectively for $\Delta z = 24\mu$m at $t=75$
ms and $\Delta z = 15\mu$m at $t=250$ ms). The rightmost inset
demonstrates instead that for small displacements the condensate
remains coherent even after several oscillations ($\Delta z = 6\mu$m
at $t=500$ ms).

%-------------------------------------------
\section{Expansion in a waveguide}
\label{sec:expansion}
%-------------------------------------------

Let us now consider the expansion of the condensate in an optical
waveguide, in the presence of disorder. In this case we will refer to a
second experiment performed at LENS \cite{fort}. Similar experiments
have also been performed by D. Cl\'ement \textit{et al.}
\cite{clement}.  The condensate is initially confined in an optical
harmonic trap of frequencies $\omega_z=2\pi\times 30$ Hz and
$\omega_\perp=2\pi\times 300$ Hz in the presence of a speckle potential
of intensity $V_0=0.2~\mu_{TF}$, 
($\mu_{TF}\simeq 87\hbar\omega_z$ 
is TF chemical potential of the condensate in the optical harmonic trap). 
The condensate is prepared in the ground state of the combined potential,
and then let expand through the waveguide by switching off the axial
trapping.

Owing to the strong radial confinement the expansion of the system can
be conveniently described by the non-polynomial Schr\"odinger equation
(NPSE) \cite{npse}, that can be written in a compact form as
\begin{equation}
i\hbar\frac{\partial}{\partial t}\varphi=
\left[-\frac{\hbar^2}{2m}\nabla_z^2+V(z)
+\frac{\hbar\omega_\perp}{2}\left(
3\sigma^2-\frac{1}{\sigma^2}\right)\right]\varphi
\end{equation}
with $\sigma^2=\sqrt{1+2aN|\varphi|^2}$.

\begin{figure}
\includegraphics[width=0.95\columnwidth,clip=]{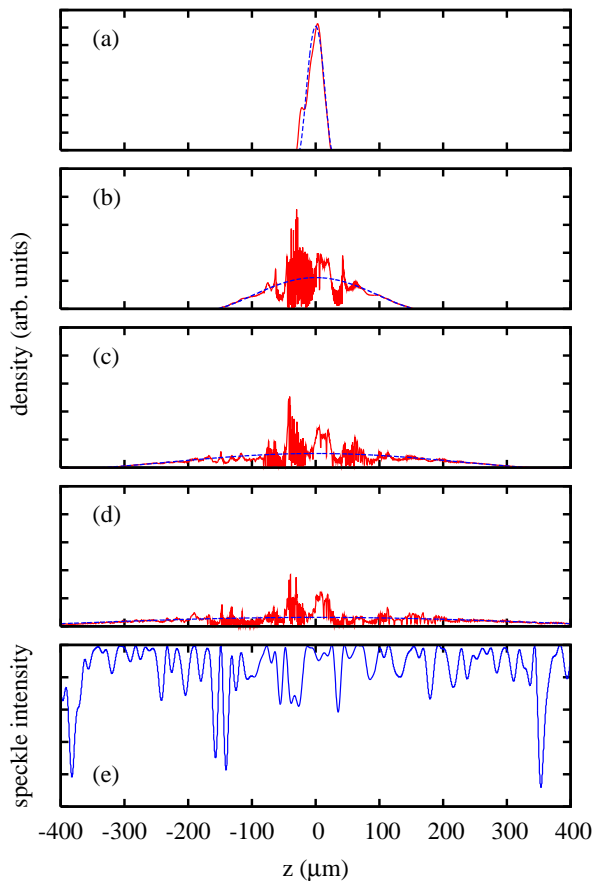}
\caption{(Color online) (a)-(d) Density profiles of the condensate
(red continuous line) during the expansion in the waveguide in the
presence of a red-detuned speckle potential (shown in frame (e)), for
different times ($t=0,\,25,\,50,\, 75$ ms, from (a) to (d)), compared
with the free expansion case (blue dashed line). Notice that the
$y$-axis scale changes from (a) to (d).}
\label{fig:exp_red}
\end{figure}

\begin{figure}
\includegraphics[width=0.95\columnwidth,clip=]{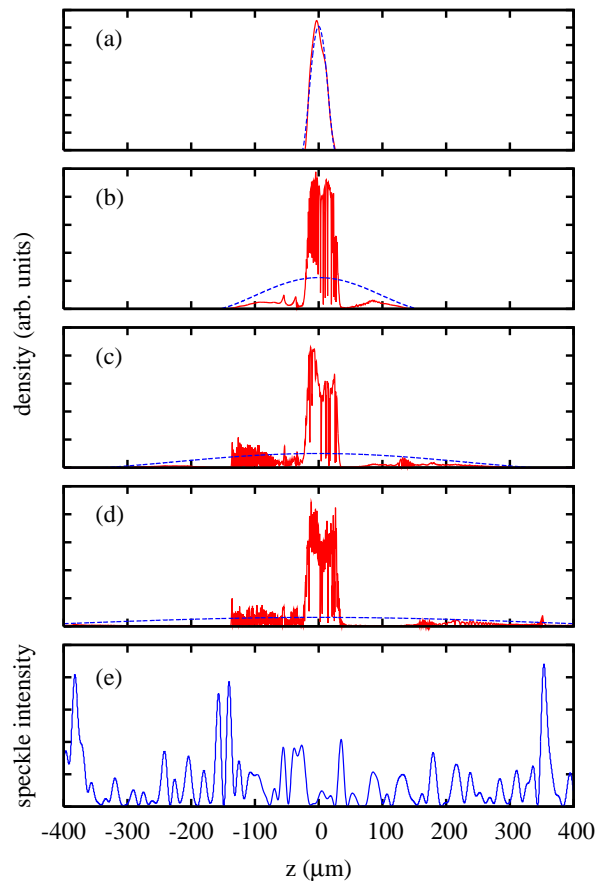}
\caption{(Color online) (a)-(d) Density profiles of the condensate
(red continuous line) during the expansion in the waveguide in the
presence of a blue-detuned speckle potential (shown in frame (e)), for
different times ($t=0,\,25,\,50,\, 75$ ms, from (a) to (d)), compared
with the free expansion case (blue dashed line). Notice that the
$y$-axis scale changes from (a) to (d).}
\label{fig:exp_blu}
\end{figure}

\begin{figure}
\includegraphics[height=\columnwidth,clip=,angle=-90]{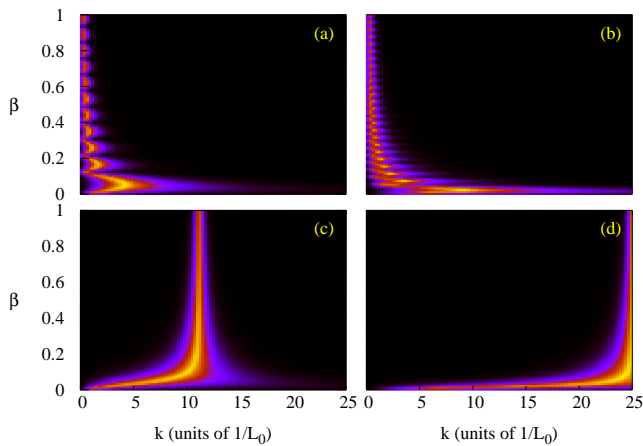}
\caption{(Color online) Density plots of the function $q(k,\alpha,\beta)$
(see text) for a potential well ((a)-(b)) and a barrier ((c)-(d)),
as a function of the incident momentum $k$ and the
length scale $\beta$ of the potential. The maximum value of $k$  
in the figures corresponds to an energy $E=U_0$.
Left and right columns refer to different potential intensities:
(a)-(c) $\alpha=0.2$, (b)-(d) $\alpha=1$. 
The plot in Figs. \ref{fig:exp_red},\ref{fig:exp_blu} must be compared
with the case $\beta=1$ in (a) and (c) respectively. 
Dark regions indicate complete reflection or transmission, light gray 
(color online) corresponds to a $50\%$ transparency.}
\label{fig:quantum}
\end{figure}

\begin{figure}
\includegraphics[width=0.95\columnwidth,clip=]{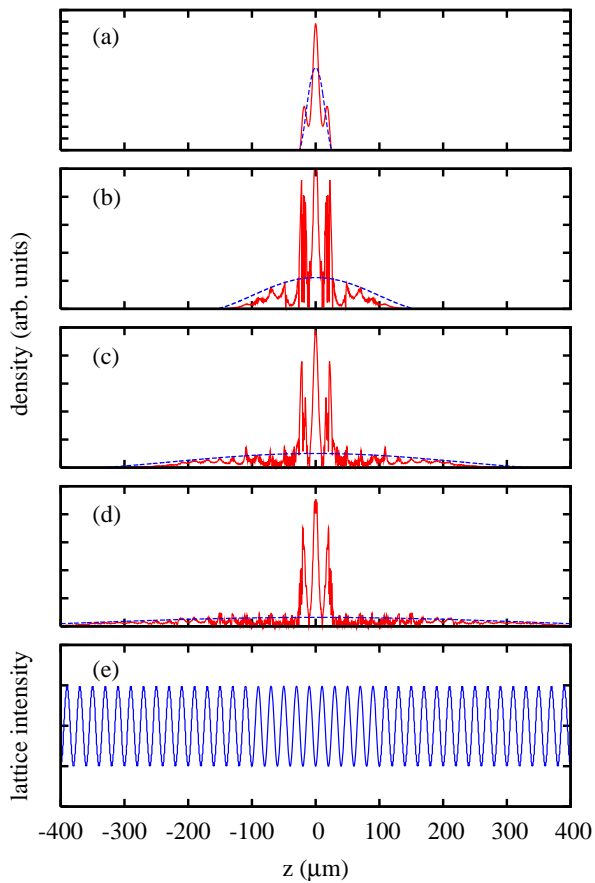}
\caption{(Color online) (a)-(d) Density profiles of the condensate
(red continuous line) during the expansion in the waveguide in the
presence of a periodic lattice (shown in frame (e)), for
different times ($t=0,\,25,\,50,\, 75$ ms, from (a) to (d)), compared
with the free expansion case (blue dashed line). Notice that the
$y$-axis scale changes from (a) to (d).}
\label{fig:exp_per}
\end{figure}

In Figs.  \ref{fig:exp_red}-\ref{fig:exp_per} we show the density
profiles of the condensate at different times during the expansion in
the waveguide, for different choices of the random potential.
For comparison we also show the corresponding profiles in case of a 
free expansion in the waveguide.

Let us discuss the figures by starting from the red-detuned speckles
(the case of Ref. \cite {fort}) in Fig.~\ref{fig:exp_red}. In this
case the dynamics is characterized by an almost free expansion of the
lateral wings of the condensate, whereas the central part remains
localized in the deepest wells of the potential.  This behavior can be
easily explained by recalling that in the TF
regime and in the absence of disorder the velocity field has a linear
dependence on $z$, $v(z,t)=z{\dot{\lambda}_z(t)}/{\lambda_z(t)}$
($\lambda_z(t)$ is a scaling parameter \cite{bec_review}), indicating
that the most energetic atoms reside at the edges of the condensate
whereas the atoms close to the center have a nearly vanishing
velocity. The presence of a weak disorder does not modify
substantially this picture. This means that the outer part of the
condensate can be sufficiently energetic to pass over the defects of the
potential expanding as in the unperturbed case, whereas the central
part remains partially localized in the initially occupied wells (see
the two density peaks in the center of the figures). A closer look to
Fig.~\ref{fig:exp_red} shows also that in the intermediate region the
density distribution shows peaks that are instead in correspondence of
the maxima of the potential as a consequence of the acceleration
acquired across the potential wells during the expansion.

In the presence of blue-detuned speckles, see
Fig.~\ref{fig:exp_blu}, the behavior is similar although in this case
the condensate may undergo a reflection from the highest
barriers that eventually stop the expansion as happens at the left
side of the particular disorder realization in the figure. Even in
this case the central part of the condensate gets localized, being
trapped by two barriers that act as a potential well in the previous
case \cite{clement}. 
We have also verified that, as one would expect, the case of a
gaussian random disorder is characterized by an intermediate behavior
between the former two, with part that is reflected by the highest barriers
and part that is localized in the central wells.

A central question is whether the observed behavior is of classical or
quantum nature. Indeed, to observe non trivial localization effects
caused by multiple interference of the condensate in the speckle
potential, the single wells/barriers should behave as quantum
reflectors \cite{nota2}.  A qualitative insight on the behavior of the random
potential can be therefore obtained by considering the case of a
single defect.  In case of the speckles a suitable model is a
sech-squared potential of the form
\begin{equation}
U(z)=\alpha U_0{\rm sech}^2\left(\frac{z}{\beta L_0}\right)
\label{eq:sech2}
\end{equation}
where $\alpha$ and $\beta$ are scaling factors for energies and
lengths respectively. The transmission coefficient of this potential
is known analytically \cite{landau}
\begin{equation}
T=\cases{\displaystyle
\frac{\sinh^2(\pi\beta\tilde{k})}{\sinh^2(\pi\beta\tilde{k})+
\cos^2\left(\displaystyle\frac{\pi}{2}\sqrt{1-4\alpha\beta^2\eta}\right)},
\; 4\alpha\beta^2\eta\le1 \cr 
\cr
\displaystyle\frac{\sinh^2(\pi\beta\tilde{k})}{\sinh^2(\pi\beta\tilde{k})
+ \cosh^2\left(\displaystyle\frac{\pi}{2}\sqrt{4\alpha\beta^2\eta-1}\right)},\,
4\alpha\beta^2\eta>1\cr}
\end{equation}
with $\eta\equiv 2mU_0 L_0^2/\hbar^2$ and 
$\tilde{k}\equiv kL_0=\sqrt{2mE}L_0/\hbar$, $E$
being the energy of the incoming wavepacket.  For convenience here we
set the energy scale to the TF chemical potential of the condensate,
$U_0=\mu_{TF}$.  The length scale instead is fixed to $L_0=\gamma
l_c$, with $\gamma\simeq2.72$, in order to match, for $\beta=1$, 
the correlation length of the potential in Eq. (\ref{eq:sech2}) with that of
the random potential. With this choice the correspondence with the
cases shown in Figs. \ref{fig:exp_red},\ref{fig:exp_blu} is for
$\alpha=0.2$ and $\beta=1$.

The ability of the above potential to act as a quantum reflector can
be suitably quantified by introducing the function
$q(k,\alpha,\beta)\equiv 2\,|\,0.5-T(k,\alpha,\beta)|-1$, that
vanishes in case of complete transmission or reflection, and equals
one for a $50\%$ transparency. In Fig. \ref{fig:quantum} we show a
density plot of $q$ as a function of $k$ and $\beta$ for two values of
$\alpha$, considering both the case of a potential well and of a
barrier (that are relevant respectively for the comparison with the
red and blue detuned speckles). The range chosen for the incident
momentum $k$ corresponds to energies up to $U_0$. The figure shows
that in case of the speckles with a correlation length as in the
experiment ($\beta=1$, $\alpha=0.2$) the range of energies where
quantum effects are evident is just a very narrow region close to the
top of the barrier or at the well border. For this value of the
correlation length even increasing the intensity of the potential (a
factor of $5$ in the figure) the situation does not change
substantially. Instead, by reducing the length scale of the disorder
($\beta\rightarrow0$) quantum effects may eventually become predominant
in a wide range of energies. This corresponds to the fact that 
the height of the single defect should
vary by a quantity at least of the order of the energy $E$ of the
incoming wave packet in a distance short compared to its de~Broglie
wavelength $\lambda_{dB}$, that is: $|dU/dz|\; \lambda_{dB}>E$.  As
discussed and experimentally demonstrated in \cite{fort}, the above
condition becomes very difficult to fulfil when the defects are
created by near-infrared light as for a speckle potential.

These considerations suggest the interpretation of the observed
localization as a classical effect due to the actual shape of the
potential. In this picture the condensate gets partially localized by the
presence of high barriers \cite{clement} or deep wells \cite{fort} in
the potential that act as single traps when the \textsl{local} chemical
potential becomes of the order of their height.  This is also
confirmed by the comparison with the case of the periodic potential,
that presents a qualitatively similar behavior as shown in
Fig.~\ref{fig:exp_per}. Even in this case the most energetic part of
the condensate expands nearly as free, whereas the bulk remains trapped
in the central wells of the potential.  The same picture holds even in
case of a single well, as discussed in \cite{fort}.

Concerning the role of interactions, we note that they introduce the
dephasing at the origin of the fast density modulations shown in the
figures, that may eventually lead to a breakdown of the superfluid
flow as discussed in Section \ref{sec:gpe}.  Their possible
contribution to localization instead is not evident. Rather they act
against localization, since they are responsible of the fast expansion
of the lateral wings (the expansion in the noninteracting case would
be much slower).

%------------------------------------------- 
\section{Discussion and conclusions}
\label{sec:conclusions}
%------------------------------------------- 

A general analysis of the effects of a weak disorder created by
speckle light on the collective modes and the expansion of an
harmonically trapped condensate has been presented by using the
Gross-Pitaevskii (GP) theory. The effects of different kinds of random
potentials and a systematic comparison with the case of a periodic
lattice with spacing of the order of the length scale of the disorder
have been also discussed.

 In the small amplitude regime the dipole and quadrupole modes are
undamped and characterized by uncorrelated frequency shifts that
depend on the particular realization of disorder. This behavior,
predicted by a perturbative sum rules approach, has been confirmed by
the direct solution of the GP equation and observed in the experiment
\cite{lye}. The theoretical analysis shows also that the average
features do not depend crucially on the particular kind of disorder,
but are however significantly different from the periodic case.

When released in a 1D waveguide the condensate may be
trapped into single wells or between barriers of the random potential,
yielding a reduced expansion. These phenomena are of
\textit{classical} nature and take place preferably near the trap
center where the less energetic atoms reside. The outer part of the
condensate instead expands almost freely, unless it encounters a high
enough (reflecting) barrier. This behavior has been observed in recent
experiments where the condensate is let expand in the presence of
potential wells \cite{fort} or barriers \cite{clement}. In the first
case the qualitative behavior is very similar to that of a periodic
system or even of a single well.

We notice that in order to observe Anderson localization or related
phenomena in a 1D waveguide one should instead have interference of multiple
\textit{quantum} reflections of matterwaves.  This regime could be
achieved by reducing the correlation length of the random potential,
but may be not a trivial task due to the diffraction limit on the size
of the defect created by light \cite{fort}.

In this respect the present analysis, besides providing useful
informations on the superfluid behavior of a condensate in the
presence of a rough surface potential, suggests that it would be
interesting to engineer other kinds of potentials by reducing the
spacing or increasing the steepness.

%------------------------------------------- 
\acknowledgments
%------------------------------------------- 

I thank L. Fallani, C. Fort, and J. E. Lye for fruitful suggestions and 
a critical reading of the manuscript, and V. Guarrera, D. S. Wiersma, and 
M. Inguscio for stimulating discussions.  This work has been supported
by the EU Contract HPRN-CT-2000-00125 and by MIUR PRIN 2003.

%-------------------------------------------
\appendix
\section{random distributions}
\label{sec:random}
%-------------------------------------------

In this section we discuss how the random distributions used in the
paper are constructed and characterized. For simplicity here we will use
dimensionless units (lengths are expressed in units of an arbitrary
scale $\xi$ whose actual value is irrelevant here).
\begin{figure}
\includegraphics[height=0.95\columnwidth,clip=,angle=-90]{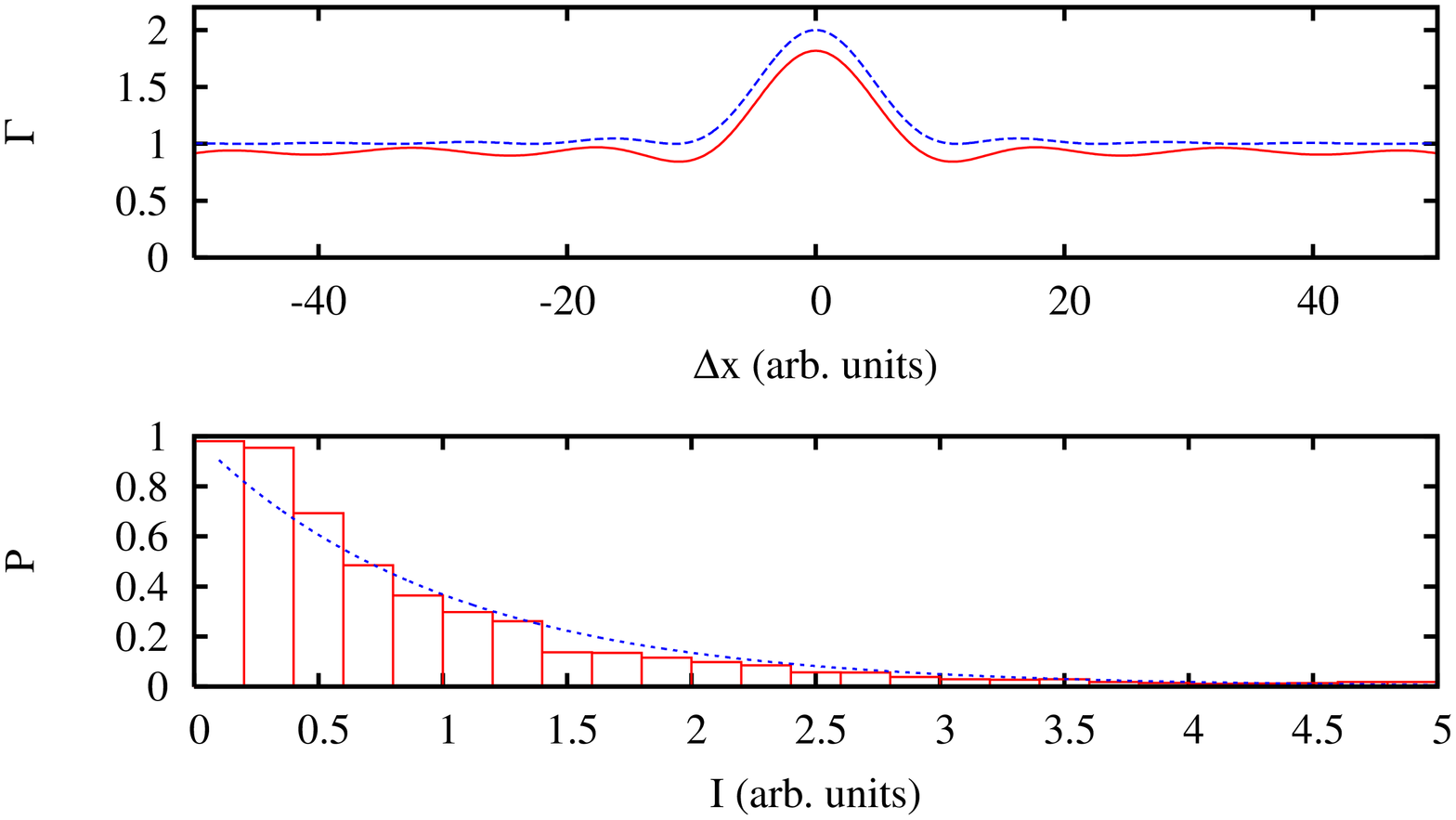}
\caption{(Color online) Continuous line (red): autocorrelation
function (top) and intensity distribution (bottom) for the speckle
potential in Fig.~\ref{fig:pot}. The (blue) dashed lines represent the
expected average values over several realizations.}
\label{fig:speckles}
%\end{figure}
%\begin{figure}
\includegraphics[height=0.95\columnwidth,clip=,angle=-90]{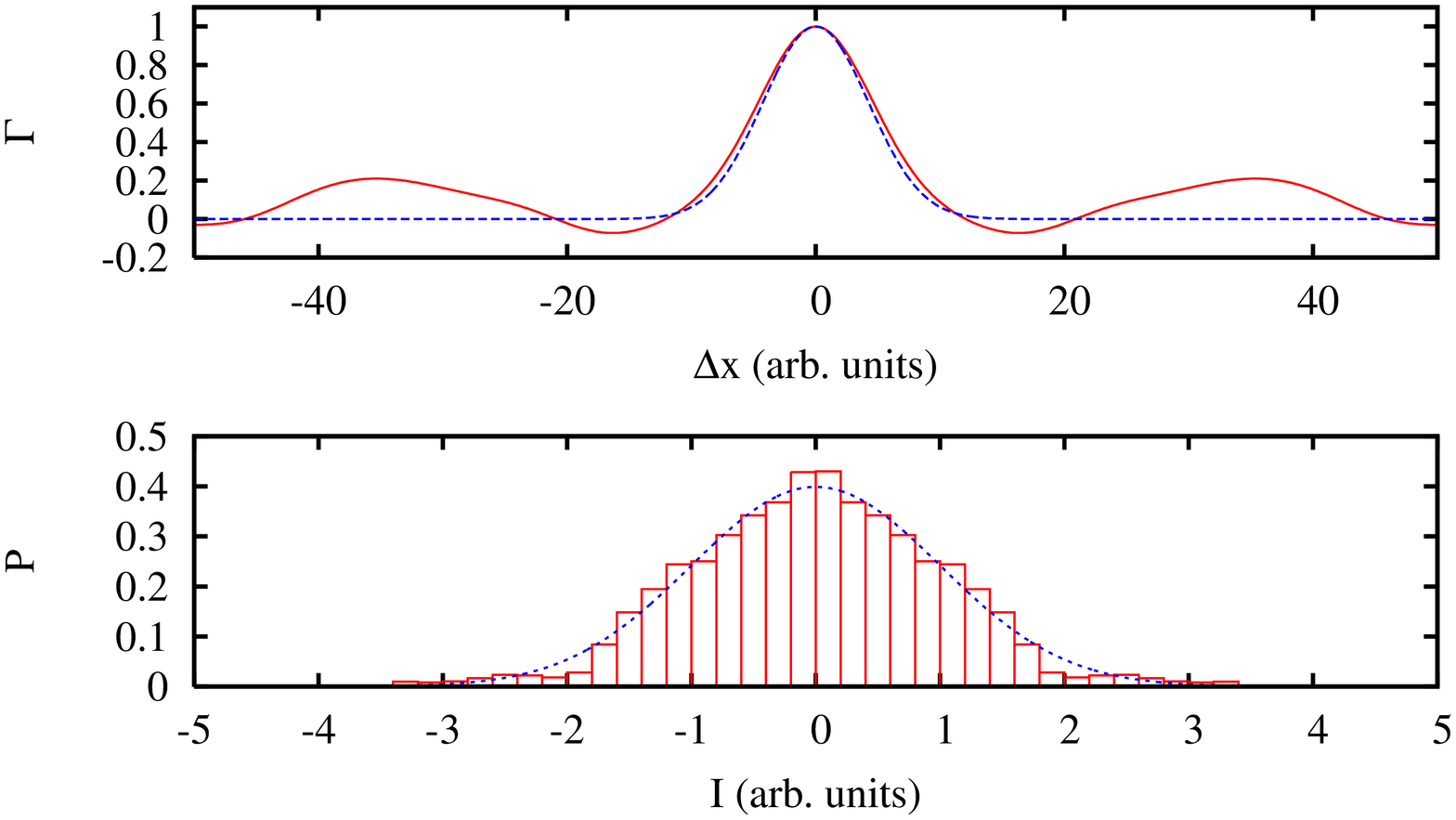}
\caption{(Color online) Autocorrelation function (top) and intensity
distribution (bottom) for the gaussian random potential in
Fig.~\ref{fig:pot}.  The (blue) dashed lines represent the expected
average values over several realizations.}
\label{fig:gaussian}
%\end{figure}
%\begin{figure}
\includegraphics[height=0.95\columnwidth,clip=,angle=-90]{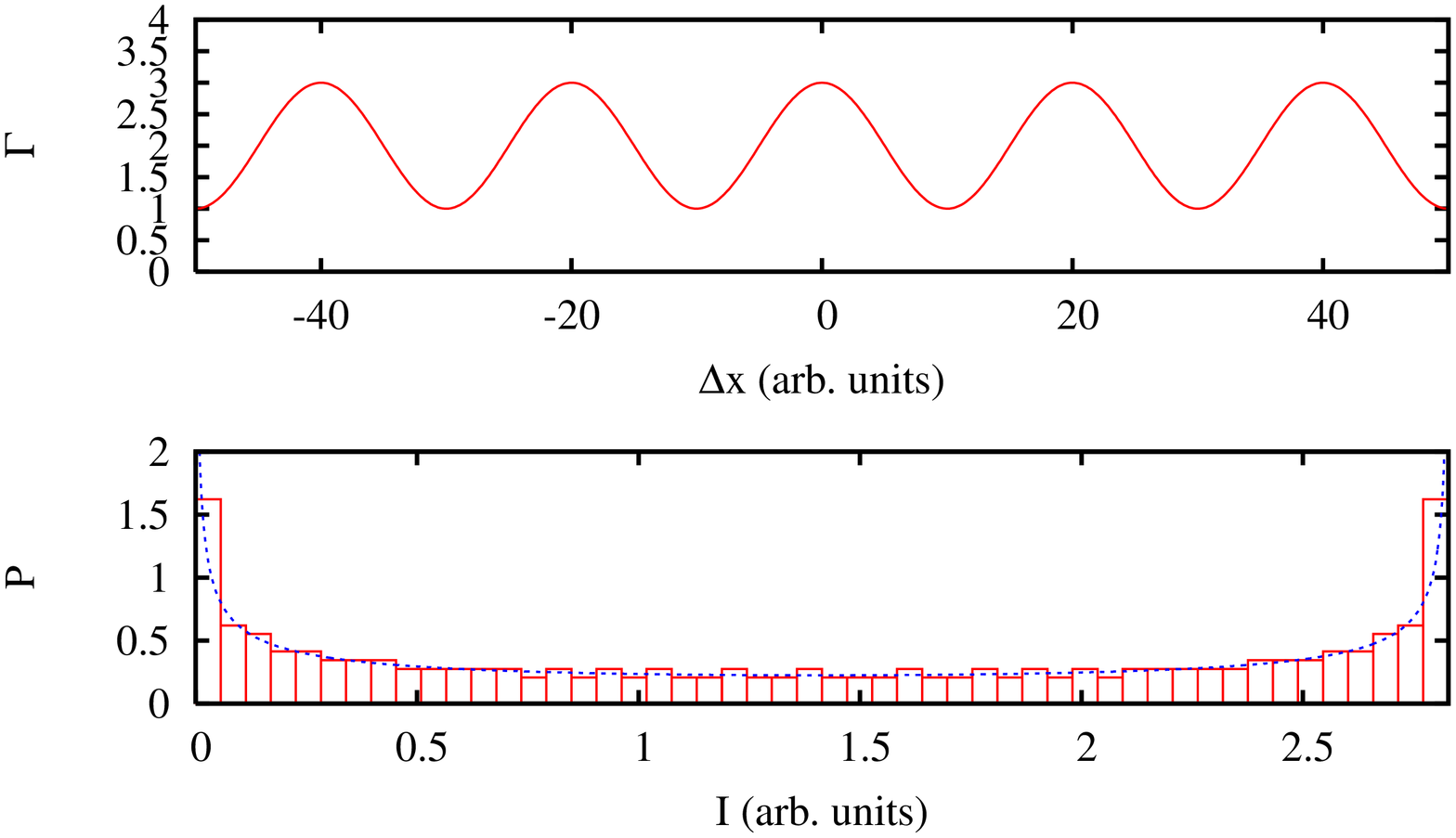}
\caption{(Color online) Autocorrelation function (top) and intensity
distribution (bottom) for the periodic potential in
Fig.~\ref{fig:pot}.The (blue) dashed lines represent the expected
average values over several realizations.}
\label{fig:periodic}
\end{figure}

Following \cite{huntley} the \textit{speckle} distribution is
constructed by starting from a random complex field $\varphi(x)$ (on a
grid) whose real and imaginary part are obtained from two independent
gaussian random distribution $\eta(x)$ with zero mean $\langle
\eta(x)\rangle$=0, unit standard deviation, and correlation function
$\langle \eta(x)\eta(y)\rangle\sim\delta(x-y)$.  The speckle intensity
field is then defined as
\begin{equation}
I(x)=|{\cal F}^{-1}\left[W(y){\cal F}[\varphi(x)]\right]|^2
\end{equation}
where the operator ${\cal F}$ indicates the Fourier transform
\begin{equation}
{\cal F}[\varphi]=\int dx \varphi(x)\textrm{e}^{2\pi i xy}
\end{equation}
and $W(y)$ indicates the aperture function

\begin{equation}
W(y)=\cases{1 & if $|y|<D/2$\cr\cr 0 & elsewhere \cr}
\end{equation}
The resulting distribution probability of the speckle intensities is
\cite{goodman}
\begin{equation}
P(I)=\textrm{e}^{-I/\langle I\rangle}/\langle I\rangle
\end{equation}
and can be further normalized to $\sigma_I=\langle I\rangle\equiv1$
(the normalized speckle distribution is indicated in the text as $v_S(x)$).
The spatial (auto)correlation is
\begin{equation}
\Gamma(\Delta x)\equiv \langle I(x)I(x+\Delta x)\rangle = 1 +
\textrm{sinc}(D \Delta x)^2
\end{equation}
(the average $\langle \cdot\cdot\rangle$ stands for an integration
over $x$ and an average over many realizations) with
$\textrm{sinc}(x)\equiv\sin(\pi x)/(\pi x)$.  The correlation
properties can be summarized by the \textit{correlation length} $l_c$
defined as the width at the half value of the maximum of $\Gamma(\Delta x)$ (in
$\Delta x=0$) with respect to the background. In case of a one
dimensional speckle distribution as that considered here $l_c$ is
related to the aperture width by $l_c=0.88/D$.

As a second source of disorder we consider a gaussian random
distribution defined by \cite{cheng}
\begin{equation}
g(x)={\cal F}^{-1}\left[\sqrt{{\cal F}[W(x)]}\eta(y)\right]
\end{equation}
where $\eta(y)$ itself is a gaussian random distribution (defined as
above) and the aperture function is $W(x)=\exp(-x^2/2\sigma^2)$.  By
using the properties of $\eta$ is then easy to demonstrate that both
the real and imaginary part of $g(x)$ are gaussian random distributions
with a correlation function $\Gamma(\Delta x)=W(\Delta x)$ and
 correlation length $l_c=2\sqrt{2\ln2}\sigma$.
Here we will consider in particular the imaginary component, $I(x)\equiv
\textrm{Im}[g(x)]$ 
(in the text the normalized distribution is indicated as $v_G(x)$).

Finally let us discuss how to chose the wavevector of the periodic
potential $v_L(x)$. In this case, the periodicity of the potential
reflects in the periodic structure of the correlation function
$\Gamma(\Delta x)\sim 1+2\cos^2(q\Delta x)$.  By restricting over a
single period the correlation length can be defined as above, and a
straightforward calculation yields $q=\pi/2l_c$. A suitable choice
to compare the effect of disorder to the case of an ordered lattice
described by the periodic potential $v_L(x)$ is therefore to require
the two potential to have the same correlation length. This seems a
reasonable choice as shown in Fig.~\ref{fig:pot}.

The autocorrelation functions and the intensity distributions of the
three potentials are shown in Figs. \ref{fig:speckles}-\ref{fig:periodic}.

%\clearpage

%-------------------------------------------


\begin{thebibliography}{}
\bibitem{nota1} Notice that the sum rules should provide an upper
bound for the actual frequencies obtained from the GP equation
\cite{bec_review}.  In our case, owing to the perturbative approach,
this is true for the dipole mode but not always for the quadrupole
mode.
\bibitem{nota2} This requirement is necessary in 1D, but not in higher
dimensions where interference can take place  between components
scattered at different angles.
\end{thebibliography}
\end{document}